\documentstyle[10pt,aaspptwo,epsf]{article}
\lefthead{W. L. Holzapfel \ea}
\righthead{A Millimeter-wave Receiver for Cosmology}

\def\D{{\Delta}}

\def\ea{{\it et al.} \,}

\def\s-z{S-Z}

\def\comp{Comptonization\,}
\def\pr{^{\prime}}
\def\2pr{^{\prime\prime}}

\def\greatsim{\mathrel{\raise.3ex\hbox{$>$\kern-.75em\lower1ex\hbox{$\sim$}}}}
\def\lesssim{\mathrel{\raise.3ex\hbox{$<$\kern-.75em\lower1ex\hbox{$\sim$}}}}
\def\ea{{\it et al.}\ }

\begin{document}

\title{The Sunyaev-Zel'dovich Infrared Experiment:\\  A 
Millimeter-wave Receiver for Cluster Cosmology}

\author{
W.~L.~Holzapfel\altaffilmark{1}, 
T.~M.~Wilbanks\altaffilmark{2},
P.~A.~R.~Ade\altaffilmark{3},
S.~E.~Church\altaffilmark{4},
M.~L.~Fischer\altaffilmark{5},\\
P.~D.~Mauskopf\altaffilmark{4,6},
D.~E.~Osgood\altaffilmark{7},
A.~E.~Lange\altaffilmark{4}}

\altaffiltext{1}{Enrico Fermi Institute, University of Chicago,
Chicago IL 60637.}
\altaffiltext{2}{Aradigm Corporation, 26219 Eden Landing Road, Hayward CA 94545.}
\altaffiltext{3}{Department of Physics, Queen Mary and Westfield
College, Mile End Road, London, E1 4NS, U.K.}
\altaffiltext{4}{Department of Physics, Math, and Astronomy, California
Institute of Technology, MS~59-33, Pasadena CA 91125.}
\altaffiltext{5}{Energy \& Environment Division, Lawerence Berkeley Laboratory,
1 Cyclotron Road, MS~90-3058, Berkeley CA 94720.}
\altaffiltext{6}{Department of Physics, University of California,
Berkeley, CA 94720.}
\altaffiltext{7}{Department of Agriculture and Resource Economics,
University of California, Berkeley CA 94720.}

\begin{abstract}
Measurements of the Sunyaev-Zel'dovich (S-Z) effect towards distant
clusters of galaxies can be used to determine the Hubble constant and
the radial component of cluster peculiar velocities.
Determination of the cluster peculiar velocity requires the separation of
the two components of the S-Z effect, which are due to the thermal and
bulk velocities of the intracluster plasma.  The two components can be
separated practically only at millimeter (mm) wavelengths.
Measurements of the S-Z effect at mm wavelengths are subject to
minimal astrophysical confusion and, therefore, provide an important 
test of results obtained at longer wavelengths.
We describe the instrument used to make the first significant detections
of the S-Z effect at millimeter wavelengths.  This instrument employs
new filter, detector, and readout technologies to produce sensitive 
measurements of differential sky brightness stable on long time
scales. These advances allow drift scan observations which 
achieve high sensitivity while minimizing common sources of systematic 
error.
\end{abstract}


\pagestyle{headings}

\section{Introduction}
\label{cintro}
The Cosmic Microwave Background (CMB) provides a uniform
back-light to the entire observable universe.  
Interactions between the CMB and intervening matter provide a 
potentially powerful probe of that matter. 
The hot intracluster (IC) medium gravitationally bound 
to rich clusters of galaxies was first discovered 
through it's X-ray {\it bremsstrahlung} emission.
In 1972, Sunyaev and Zel'dovich (1972) argued that 
Compton scattering by this diffuse plasma
should produce a measurable distortion of the
spectrum of the CMB viewed through the cluster.  
This has come to be known as the Sunyaev-Zel'dovich 
(S-Z) effect.

The combination of the S-Z effect and X-ray emission provides 
an extremely powerful probe of the IC medium.
Measurements of the S-Z surface brightness can be combined
with the X-ray surface brightness and spectra to determine
the Hubble expansion parameter (\cite{Caval}; \cite{BHA}).
The combination of the S-Z and X-ray spectra 
can be used to determine the radial component
of the cluster peculiar velocity (\cite{Rephaeli91}; \cite{Holzapfelb}).

Due to its low surface brightness, observation of the S-Z effect is 
difficult, for reasons which vary with
measurement frequency, angular scale, and observing platform.  
After
many years of discrepant results, reliable detections are being
made at cm wavelengths using single dish radio telescopes
(\cite{Myers}; \cite{Herbig}; \cite{BHA}), 
and close-packed radio interferometers (\cite{Jones}; \cite{Carlstrom}).  
Although there is strong
motivation, as described below, to obtain measurements at millimeter
wavelengths, these have proven even more difficult.  
This paper
describes the instrument and method used by 
Wilbanks \ea (1994) to
make the first significant detections of the S-Z effect at millimeter
wavelengths.

\subsection{The Sunyaev-Zel'dovich Effect}
\label{sszsci}
There are two velocity components of the scattering 
free electrons which produce the two distinct aspects of the
S-Z effect.
The thermal component of the S-Z effect results from the thermal motion 
of the 
electrons in the IC plasma and the kinematic component results from 
the bulk motion of the IC plasma relative to the CMB rest frame.
The two components can be treated as independent and
their brightnesses can be added linearly.

\begin{figure}[ht]
\plotone{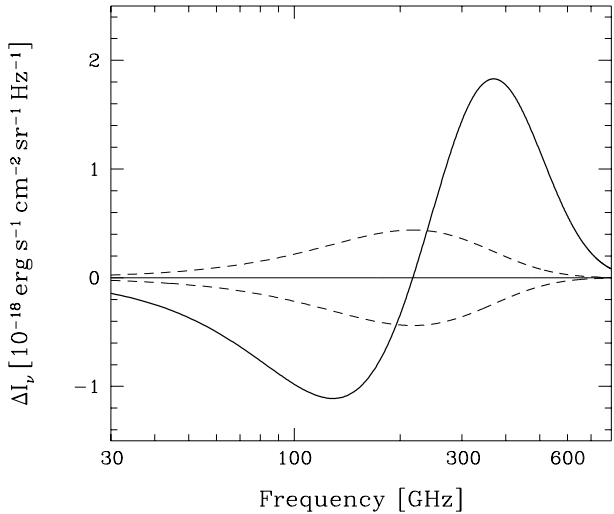}
\caption[Spectra of SZ effects]
{The spectra of the two components of the S-Z effect.
The solid line shows the change in brightness, compared to the
unperturbed CMB, of a $y = 10^{-4}$ thermal component plotted as a
function of frequency.
The dashed lines show the same for the kinematic component
corresponding to $\tau = .01$ and $v_r=\pm 1000\,{\rm kms}^{-1}$
(positive and negative curves, respectively).}
\label{SZspec}
\end{figure}

The Compton scattering, which produces the S-Z effect, preserves 
the number of photons, but on average increases their energy.
Relative to the unperturbed CMB spectrum, the thermal component produces a
decrement in brightness at low frequency and an increment in
brightness at high frequency. 
In the nonrelativistic limit, the CMB intensity change due to 
the thermal component of the S-Z effect
is given by (\cite{ZS68})
\begin{equation}
\D I_{nr} = I_{0}\, y\, g(x)\, ,
\end{equation}
where $x=h\nu/kT_{0}$, where $T_{0}$ is the CMB temperature, and
\begin{equation}
I_{0} \equiv \frac{2(kT_{0})^3}{(hc)^2}\,.
\end{equation}
The spectrum of the thermal S-Z is given by the function
\begin{equation}
g(x) = {x^4e^x \over \left(e^x -1\right)^2} \left[{x \left(e^x +1\right)\over
e^x -1} -4\right]\,, 
\label{nreqn}
\end{equation}
which vanishes at $x_0=3.83$ ($\nu_0=217\,$GHz) for $T_0=2.726\,$K.
The amplitude of the effect is given by the \comp\ parameter,
\begin{equation}
y = \int \left(\frac{kT_e}{mc^2}\right) n_e \sigma _T dl\,,
\label{y}
\end{equation}
where $n_e$ and $T_e$ are the electron density and temperature,
$\sigma _T$ is the Thomson cross section, and the integral is over a line
of sight through the cluster.
The change in intensity due to the thermal S-Z effect is plotted 
in Fig.~\ref{SZspec} for $y = 10^{-4}$, 
a typical value for a rich cluster.
Relativistic corrections to eq.~\ref{nreqn} are significant 
for typical values of electron temperature (\cite{Rephaeli95a}) and must 
be taken into account in order to accurately determine
the Hubble Constant and cluster peculiar velocities.  

The kinematic component of the S-Z effect is due to the bulk motion of
the IC plasma relative to the CMB rest frame.  
The scattered photons experience a Doppler shift 
dependent on the angle of their scattering relative to the bulk velocity.
The change in CMB brightness is given by
\begin{equation}
\D I_{K} = I_{0} \, h(x) \int
\frac{\vec{v_p} \cdot \hat{r}}{c} \, n_{e} \sigma_T dl \,,
\label{ISZK}
\end{equation}
where $v_r = {\vec v_p}\cdot{\hat{r}}$ is the component of the bulk 
velocity along the line of sight toward the observer.
The spectrum is described by $h(x)= x^4e^x/(e^x-1)^2$, equivalent to that 
for a change in the CMB temperature.
The change in the CMB intensity due to the kinematic
component of the S-Z effect is plotted in Fig.~\ref{SZspec} for
\mbox{$\tau = 10^{-2}$} and 
\mbox{$v_r = \pm 10^{3}~{\rm km} {\rm s}^{-1}$}.

\subsection{Measurements at Millimeter Wavelengths}
\begin{figure}[ht]
\plotone{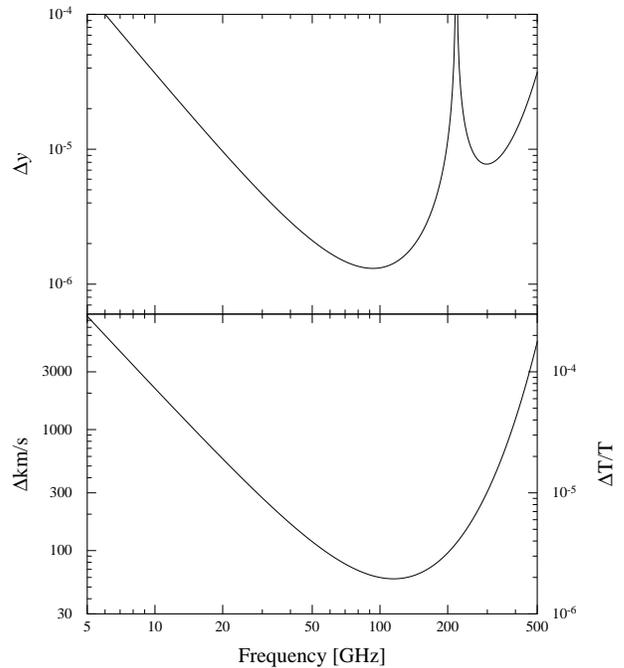}
\caption[]
{The $1\sigma$ confusion limits to measurements of the thermal (upper) and
kinematic (lower) S-Z effect on arcminute scales (from \cite{FL})
as a function of frequency.  The estimates include confusion from
radio galaxies, IR cirrus, and infrared galaxies and assume that the
S-Z source is well-matched to the beam.}
\label{SZconf}
\end{figure}

The distinct spectra of the two components of the S-Z effect
allow their separation at millimeter wavelengths.
In Fig.~\ref{SZconf} we show the expected astrophysical confusion 
from galactic and extra-galactic sources; there is a distinct
minimum at millimeter wavelengths.
On arcminute scales, which correspond to the size of distant
clusters of galaxies, known sources of astrophysical foreground 
confusion will 
limit the precision of millimeter-wave measurements of $y$ and 
$v_r$ to $\sim 10^{-6}$ and $\sim 100\,{\rm kms}^{-1}$, respectively.

Primary anisotropies of the CMB also contribute to the uncertainty
of S-Z effect measurements.
The spectrum of these distortions is identical to that of the
kinematic component of the S-Z effect and, therefore, they 
limit the accuracy with which cluster peculiar 
velocities can be determined.
The predicted amplitude of these anisotropies is a strong function of 
angular scale and cosmological model.
The confusion in peculiar velocity for a cluster of optical depth
$\tau$, due to primary anisotropies of amplitude $\D T/T$, is 
\begin{equation}
\D v_r = 100\, \left(\frac{10^{-2}}{\tau}\right)\,
\left(\frac{\D T/T}{3 \times 10^{-6}}\right)\, {\rm kms}^{-1}.
\end{equation}

Previous attempts to measure the S-Z effect at millimeter wavelengths
(\cite{Meyer}, \cite{Rea}, \cite{Chase}) have lacked the
sensitivity and/or the freedom from systematic effects necessary to
produce a reliable detection.  Instrumental offsets and response to
off-axis sources must be carefully controlled to allow accurate
measurement of this low surface brightness effect (\cite{Chase},
\cite{Uson}).  In this paper we describe a high sensitivity bolometric 
receiver that uses a drift-scan observation strategy to eliminate 
the dominant sources of systematic error.
\section{Instrument}
\label{cinst}
The Sunyaev-Zel'dovich Infrared Experiment\\  (SuZIE) is a bolometric
array receiver for use on large aperture millimeter-wave telescopes.  
It is designed to separate and accurately measure both components of the
S-Z effect in clusters of galaxies at moderate to high redshift 
($z\greatsim 0.1$).
The instrument can be configured to observe in three mm-wave 
atmospheric windows. 
A novel read-out scheme produces stable difference signals between
elements of the array, allowing drift-scan observations on time
scales of minutes.
\subsection{Optics}
\label{soptics}
SuZIE has been used at the Caltech Submillimeter Observatory (CSO) for
all observations thus far.  The telescope's location at the summit of
Mauna Kea provides excellent observing, with a zenith column density
of precipitable water vapor $\le 2\,$mm on $> 60\%$ of nights.

\begin{figure}[ht]
\plotone{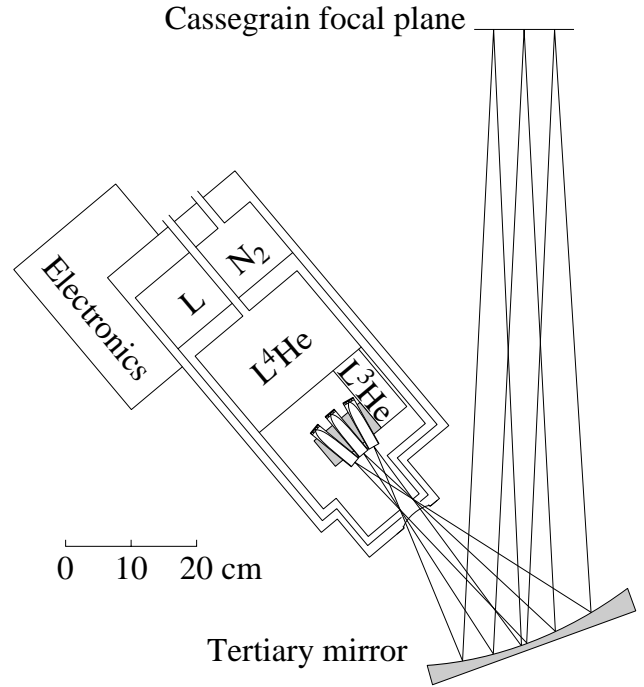}
\caption[SuZIE coupling to the Caltech Submillimeter Observatory]
{Coupling of the SuZIE receiver to the CSO using a fixed tertiary
mirror.  The cryostat can be rotated around its optical axis to
keep the array aligned with the direction of sky drift.}
\label{Oscheme}
\end{figure}

\begin{figure}[ht]
\plotone{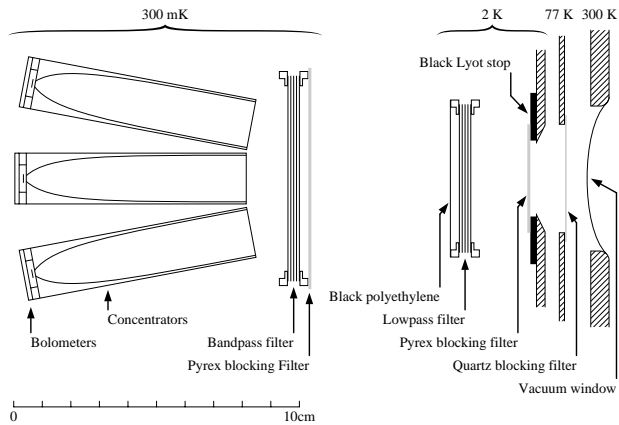}
\caption[SuZIE optics]
{Configuration of the filters and optics
inside the L$^4$He cryostat.}
\label{newoptics}
\end{figure}

The receiver optics, shown in Fig.~\ref{Oscheme} and
Fig.~\ref{newoptics}, are designed to couple the detectors to the
telescope with minimal background loading and to maximize the
overlap of the portion of the primary mirror viewed by each detector
in the array.  
The loading is minimized by using as few warm elements
as possible and by underfilling these elements to reduce spill-over.

The six detectors are mounted in low-loss cavities fed by a $2
\times 3$ array of parabolic concentrators.  The $0.20~{\rm cm}$
diameter exit apertures of the concentrators are small compared
to the absorbers, making the bolometers the dominant source of
loss in the cavities.  The $1.63\,$cm diameter entrance
apertures of the cones are nearly close packed on the tertiary
focal surface.
In time-reversed sense, the $f/2.5$ concentrators 
over-illuminate a $2\,$K, $2.5\,$cm diameter Lyot
stop placed at an image of the primary mirror.  The Lyot stop is the
throughput limiting element in the system.  It is sized so that the
$10.4\,$m diameter primary mirror is underfilled by $50\%$ in
area.

Outside the Lyot stop, the beam is clear of the other apertures in 
the optical path.  It leaves the cryostat through a $50\,\mu$m thick
polypropylene window and is incident on a $33\,$cm diameter, $40^{\circ}$
off-axis, underfilled ellipsoidal tertiary mirror.  The measured
emissivity of the $50\,\mu$m thick polypropylene window is $0.025\%$ at
$\nu = 300\,$GHz, in good agreement with the published properties
of this material (\cite{Afsar}).  
The tertiary mirror reimages the entrance apertures of the parabolic 
concentrators to the Cassegrain focus of the CSO.
The $f/\#$ of the beam emerging from the cryostat is
determined by the Lyot stop, it is $f/4$ from the receiver to the
tertiary and $f/10$ from the tertiary to the secondary.

The tertiary mirror is designed to simultaneously optimize the
image quality across the focal plane and the overlap of the beams
on the primary mirror.
We have mapped the illumination of the primary mirror with a 
mobile $80\,{\rm cm}^2$ chopping ambient load.
The measured illumination pattern at $217\,$GHz falls off by $-60\,$dB
(the limit of our signal to noise) at a radius of $4.0\,$m.
The degree of overlap of the beams on the primary determines the 
ability of the detector differences to subtract near field atmospheric 
emission.
By subtracting the illumination patterns measured for the different 
array elements, we estimate the RMS mismatch of the beams on the 
primary mirror.
In the row of detectors which lies closer to the optical axis,
the mismatch is $\sim 4\%$ and $7\%$ for elements 
separated by $2.2^{\prime}$ and $4.6^{\prime}$ respectively.
In the row which lies further off axis after the 
tertiary mirror, the mismatch is $\sim 6\%$ and $10\%$ 
for elements separated by $2.2^{\prime}$ and 
$4.6^{\prime}$ respectively.
This mismatch limits our ability to remove near field differential
atmospheric emission.

\subsection{Filters}
\label{sfilters}
The passbands are matched to atmospheric transmission windows  
in order to maximize the sensitivity of the system in the presence
of atmospheric noise.
The $142$ and $268\,$GHz bands maximize the ratio of 
S-Z thermal component signal to the combination of 
atmosphere and detector noise.
The thermal component null is near the frequency at 
at which the kinematic component is brightest and 
falls within an excellent atmospheric window.
By observing in this window, the $217\,$GHz band achieves high 
sensitivity to the kinematic S-Z component while minimizing the 
contribution of the typically much larger S-Z thermal component.
The width of the band is determined by maximizing the 
ratio of the S-Z kinematic component signal to the sum of atmosphere
and detector noise.
 
The passbands are
defined by a combination of resonant-mesh band-pass filters and
low-pass blocking filters configured as shown in
Fig.~\ref{newoptics}.  The blocking filters serve the additional
function of minimizing the radiative heat load on the cryogenic
systems from the warm optics.

A disk of fused quartz mounted on the LN$_2$ cooled shield absorbs
radiation at frequencies between $4800$ and $6\times10^4\,$GHz, blocking
most of the power from ambient-temperature sources.  The thickness of
the quartz is chosen to minimize the in-band reflection losses.  A
$0.54\,$mm thickness is used for observations at $142$ and
$268\,$GHz and a $0.345\,$mm thick disk is used for observations at
$217\,$GHz.

A Pyrex (Owens-Corning) glass disk mounted on the $2\,$K shield
directly behind the Lyot stop absorbs between $900$ and $6\times10^4\,$GHz,
blocking most of the power emitted by the $\sim 77\,$K LN$_2$
cooled portions of the cryostat (including emission from the quartz
filter) and most of the residual power from the atmosphere and other
sources outside the cryostat.  A $1.05\,$mm thickness is used for
observations at $142$ and $268\,$GHz, and a $0.67\,$mm thick disk is
used for $217\,$GHz observations.

A $150\,\mu$m thick layer of polyethylene impregnated with
carbon-black, LiF, and diamond dust is mounted on the $2\,$K
stage, immediately behind a low-pass multi-element resonant mesh
filter.  The black polyethylene has transmission $< 10^{-3}$ at $\nu >
9000\,$GHz.  Together, the quartz, Pyrex, and polyethylene filters
prevent any significant thermal load from reaching any of the
components cooled to $300\,$mK by the $^3$He refrigerator.

Each spectral band is defined by the response of
a resonant mesh band-pass filter and a matched resonant mesh
low-pass filter which blocks leaks at frequencies higher than the pass-band.  
The band-pass filter is located directly in
front of the concentrators, the low-pass filter is located midway
between the Lyot stop and the concentrators. The low-pass filter is 
slightly tipped to prevent resonance between the two metal mesh
filters. For the $142\,$GHz observations, we use a low-pass filter with 
a cut-off of $150\,$GHz. 
In both the $217$ and $268\,$GHz observations,
a low-pass with a cut-off of $300\,$GHz is used.
A second piece of Pyrex glass is placed in front of the band
pass filter to further damp any resonance between the filters
and to provide mechanical protection for the delicate
mesh filter during work on the instrument.

\begin{figure}[ht]
\plotone{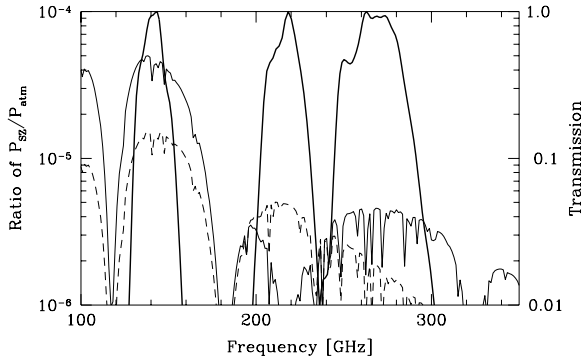}
\caption[System spectral transmission optimization]
{Ratio of S-Z effect power to atmospheric power, for 1 mm
column density of precipitable water vapor.
The fine solid line
represents this ratio for the thermal component ($y=10^{-4}$),
while the dashed one
represents the ratio for the kinematic component
$(\tau=.01,\ v_r=10^3{\rm kms}^{-1})$.  The heavy lines show
the transmission of the complete system, measured on a Fourier
transform spectrometer, for each of three frequency bands.}
\label{sysbands}
\end{figure}

The total spectral response, $f(\nu)$, measured using a Fourier
transform
spectrometer (FTS) is shown in Fig.~\ref{sysbands} for each of 
the three bands.
We define the central frequency of the band, when observing a 
source of intensity $I_{\nu}$, as 
\begin{equation}
\nu_0 = \frac{\int I_{\nu}\, f(\nu)\, \nu\, d\nu}{\int I_{\nu} \, f(\nu)\, d\nu}\,.
\end{equation} 
The width of the band is then
\begin{equation}
\D \nu = \frac{\int I_{\nu} f(\nu) d\nu}{I_{\nu_0}}\,.
\end{equation} 
We have computed the central frequency and width for 
observations of several sources.
Because the bands are narrow, the changes in the effective
central frequencies are small. 
For example, the difference in the 
central frequency when the bands are convolved with  
$300\,$K and $3\,$K blackbodies is $<0.3\%$. 
In Table~\ref{bandtab}, we list the central frequency and width of all
six array elements for each band, where $I_{\nu}$ is 
taken to be a constant.

\begin{table*}[htbp]
\begin{center}
\begin{tabular}{ccccccc}
Pass-band & \multicolumn{2}{c}{$142\,$GHz} & \multicolumn{2}{c}{$217\,$GHz} & \multicolumn{2}{c}{$269\,$GHz}\\ 
Channel & center & width & center & width &center & width\\ \tableline
$1$ & $141.15$ & $13.04$ & $216.62$ & $12.98$ & $267.95$ & $23.76$\\
$2$ & $140.67$ & $12.24$ & $215.05$ & $15.36$ & $265.80$ & $23.21$\\
$3$ & $141.69$ & $12.10$ & $215.69$ & $17.26$ & $269.40$ & $30.09$\\
$4$ & $143.35$ & $13.22$ & $218.98$ & $16.11$ & $269.44$ & $25.33$\\
$5$ & $140.73$ & $12.62$ & $215.82$ & $14.36$ & $268.71$ & $23.71$\\
$6$ & $141.93$ & $12.81$ & $218.09$ & $13.47$ & $269.97$ & $27.86$\\ \tableline
Average & $141.59$ & $12.67$ & $216.71$ & $14.96$ & $268.54$ & $25.66$\\
$\sigma$ & $1.00$ & $0.44$ & $1.53$ & $1.64$ & $1.51$ & $2.76$\\
\end{tabular}
\end{center}
\caption[]
{Center frequency and width [GHz] of the filter bands for
each detector element.
At the bottom of the columns we list the averages for each
band and the standard deviation of the
individual array elements about the mean.}
\label{bandtab}
\end{table*}

The beams from the elements on the outside corners of 
the array (1, 3, 4, and 6 in Table~\ref{bandtab})
are incident on the bandpass filter at an angle $\approx 11^{\circ}$
from normal, while the inside elements (2 and 5) are incident
with an angle of $\approx 5^{\circ}$. 
This gives rise to systematically higher central frequencies  
for the outside elements of the array.
The difference in central frequency between the inside and outside
elements is $\sim.9\%$, $.9\%$, and $.7\%$
of the central frequency in the $142$, $217$ and $268\,$GHz bands.  
Although small, the differences in the spectral response of
the array elements must be taken into account in order to
determine accurate peculiar velocities (\cite{Holzapfelb}).

In Table~\ref{oetab}, we list the optical efficiency at the peak 
transmission for each of the six detector elements in  
the three filter pass-bands.
The optical efficiencies are computed by comparing the measured
difference in power absorbed by the detectors when
viewing two fixed temperatures with that computed by integrating
the computed difference in brightness over the measured 
spectral band,
\begin{equation} 
\epsilon = \frac{(P_{T2}-P_{T1})}{A\Omega \int f(\nu)\, (I_{T2}-I_{T1})\, d\nu}\,.
\end{equation} 
$f(\nu)$ is the measured filter transmission (normalized 
to one at its peak) and $A\Omega = .104\,{\rm cm}^2{\rm sr}$ 
is the maximum geometric throughput allowed by the Lyot stop.
Due to the tapered illumination pattern of the parabolic 
concentrators, the Lyot stop is not uniformly illuminated. 
Therefore, the assumed throughput is an overestimate and the 
resultant peak optical efficiencies are lower limits to
the true values.
The average optical efficiencies at the peak of the bands 
are $37\%$, $42\%$, and $16\%$ for the 
$142$, $217$, and $268\,$GHz bands, respectively.
The lower optical efficiency of the $268\,$GHz band is due 
largely to the absorption of the Pyrex blocking filters.

\begin{table}[htbp]
\begin{center}
\begin{tabular}{cccc}
& \multicolumn{3}{c}{Filter Band}\\
Channel & $142\,$GHz & $217\,$GHz & $269\,$GHz\\ \tableline
$1$ & $29.2\%$ & $46.0\%$ & $13.7\%$\\
$2$ & $33.0\%$ & $36.3\%$ & $13.0\%$\\
$3$ & $37.2\%$ & $35.2\%$ &  $9.3\%$\\
$4$ & $35.4\%$ & $38.5\%$ & $15.2\%$\\
$5$ & $37.9\%$ & $47.8\%$ & $15.2\%$\\
$6$ & $32.8\%$ & $49.3\%$ & $13.6\%$\\ \tableline
Average & $34.2\%$ & $42.2\%$ & $13.3\%$\\
$\sigma$ & $3.2\%$ & $6.3\%$ & $2.2\%$\\
\end{tabular}
\end{center}
\caption[]
{Optical efficiency of array elements in each of the filter bands.
At the bottom of the columns we list the averages for each
band and the standard deviation of the
individual array elements about the mean.}
\label{oetab}
\end{table}

Bolometers are inherently broad-band detectors; the spectral
response of the instrument is defined by the filters which
precede the detectors.
Care must be taken to limit the response of the system
to out of band radiation from sources with rising
spectra.
For each of the three frequency bands,
we have determined the ratio of in band to out of band power 
when viewing the atmosphere, galactic dust, and a $300\,$K black-body.
The model atmosphere is computed assuming a zenith optical depth, 
$\tau(225\,{\rm GHz}) = 0.04$,
and the temperature of the atmosphere to be $260\,$K. 
The dust is assumed to have a temperature of $30\,$K and an
emissivity $\epsilon \propto \nu^{1.5}$.

The results of our tests for out of band response are summarized in
Table~\ref{leaktab}.  For the $142$ and $268\,$GHz configurations we
computed the out-of-band signal by integrating the source brightness
over the spectral response of the system.
The spectral response was measured from ($0 - 3000\,$GHz),
and computed at higher frequencies using 
transmission data obtained in our laboratory and the literature
(\cite{Halpern}; \cite{Bock}; \cite{redbook}).

For the $217\,$GHz configuration, we did not measure the 
filter transmission to high frequency using the FTS.
Instead, we measured the integrated system
response to a quartz encapsulated mercury vapor lamp that was 
low-pass filtered at $1500\,$GHz, with and without the band 
blocked by a $300\,$GHz thick-grill high-pass filter.  
The limit on the ratio of out of band to in band power is
$R_{MVL}=P(300-1500\,{\rm GHz})/P_{band} < 1/500$.
The computed response from frequencies $>1500\,$GHz is
negligible in comparison.
In order to determine the response of our system to
sources other than the Mercury vapor lamp, we must take the different 
spectra of these sources into account.
We treat the effective emissivity of the lamp as constant, although it
actually rises with frequency due to its hot quartz envelope.
For $30\,$K dust we assume the worst case: that all of the out of 
band power contribution comes from $1500\,$GHz. 
This is very conservative because the Pyrex transmission is computed
to be $\sim 10^{-6}$ at this frequency.
The ratio of in to out of band power from $30\,$K dust is then 
the ratio of in band to out of band power from the Mercury vapor lamp 
scaled by the frequency dependence of the dust emission, 
$R_{MVL}\times \left(1500/217\right)^{1.5}$.
We can perform a similar calculation for the atmosphere by assuming that 
the atmosphere outside the band is totally opaque.
The ratio of in band to out of band power is then 
${R_{MVL}}/{\left(1-e^{-\tau_{217}}\right)}$ where $\tau_{217}$ is the
opacity of the atmosphere in the $217\,$GHz band.

\begin{table*}[htbp]
\begin{center}
\begin{tabular}{cccc}
Pass-band & \multicolumn{3}{c}{Source} \\
\ [GHz] & $300\,$K & $30\,$K, $\epsilon \propto \nu^{1.5}$ & $260\,$K Atmosphere\\ \tableline
$142$\tablenotemark{a} & $< 0.004$ & $< 0.024$ & $< 0.060$ \\
$217$\tablenotemark{b} & $< 0.002$ & $< 0.036$ & $< 0.057$ \\
$269$\tablenotemark{a} & $< 0.013$ & $< 0.035$ & $< 0.062$ \\
\end{tabular}
\end{center}
\caption[]
{Ratio of Out-of-band to in-band response for thermal sources}
 
\tablenotetext{a}{Values determined by combining measured pass-band
spectral response and calculated dielectric filter blocking (see text).}
\tablenotetext{b}{Values determined by combining  measured out-of-band
response and calculated dielectric filter blocking (see text).}
 
\label{leaktab}
\end{table*}

By measuring the system response as a function of zenith angle, 
we have separated the contributions to the total optical loading 
from the telescope and atmosphere. 
The results are calibrated in terms of brightness temperature 
from observations of $77$ and $300\,$K loads.
In Table~\ref{Tatm} we list the telescope and atmosphere brightness 
temperatures for each array element and frequency band.
We have computed the brightness temperature of the atmosphere 
using a model incorporating FTS measurements of the 
atmosphere at the CSO (\cite{Lis}). 
The model atmosphere is assumed to be $~\sim 260\,$K and the
amount of precipitable water vapor is adjusted to match the 
measured zenith opacity, $\tau(225\,{\rm GHz})$, at the time of the 
measurements.
The results are in reasonable agreement with the model; 
a more detailed comparison would 
require information about the thermal structure of the atmosphere.
The brightness temperature of the telescope was found to be
$\sim 30\,$K for all three spectral bands.
This number is consistent with the results of other instruments used 
at the CSO; we attribute the bulk of the telescope emissivity to 
scattering from the secondary support structure.

\subsection{Detectors}
\label{sdetectors}
Bolometric detectors offer the highest sensitivity to continuum
radiation at millimeter wavelengths.  
Table~\ref{bolosens} gives a summary of the background power and
background photon noise limit (BLIP) for each of the 3 SuZIE
passbands.
Sensitivity at
or near the background limit can be achieved using composite
bolometers cooled to $300\,$mK (\cite{Alsop}).

\begin{figure}[ht]
\plotone{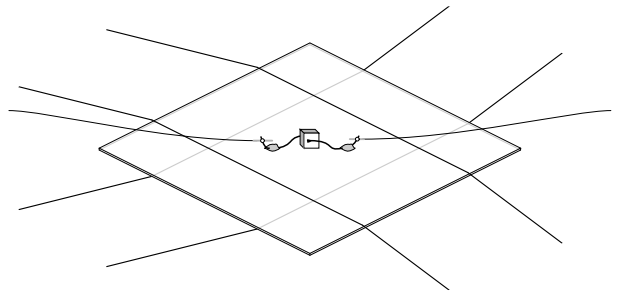}
\caption[Composite bolometer]
{A composite bolometer used in the SuZIE instrument.  The thermistor
with wedge-bonded gold leads is secured to the sapphire substrate with
two blobs of epoxy.  The NbTi electrical leads have copper-clad ends,
which are soldered to the gold wires.  The substrate is suspended by
four nylon threads.}
\label{bolo}
\end{figure}

Each of the SuZIE bolometers (Fig.~\ref{bolo}) consists of a 
thermistor mounted on a $4\, {\rm mm}^2 \times 25 \, \mu {\rm m}$ 
sapphire (crystalline ${\rm Al}_2{\rm O}_3$) absorber. 
A coating of bismuth approximately
$1000\,$\AA\ thick is evaporated on one side of the absorber.  The
thickness of the bismuth is chosen to maximize the broad-band
absorption of light incident on the bolometer from the sapphire side.
Nylon threads suspend
the absorber adjacent to the exit aperture of the concentrator cone,
with the bare side towards the cone.

\begin{table*}
\begin{center}
\begin{tabular}{ccccc}
Pass-band & $\tau(225\,{\rm GHz})$ & Telescope & Atmosphere & Model Atmosphere  \\
\ [GHz] & & \multicolumn{3}{c}{Brightness Temperature [K]} \\ \tableline
$142$ & $.045$ & $29.6\pm4.0$ & $6.5\pm1.3$ & $5.51\pm0.04$ \\
$217$ & $.048$ & $26.1\pm2.2$ & $8.4\pm0.5$ & $11.68\pm0.14$\\
$269$ & $.031$ & $28.4\pm1.0$ & $10.8\pm0.3$ & $14.06\pm0.57$ \\
\end{tabular}
\end{center}
\caption[]
{Telescope and Atmospheric temperature computed from sky dips.
The uncertainty in the measured telescope and atmosphere temperatures
are the standard deviations of the results for each array element
about their mean.
The uncertainty in the model atmosphere temperature is the standard
deviation of the calculated power in the six array elements from
the integrals of the model atmosphere
(computed for $\tau(225\,\rm{GHz})$) over the measured spectral bands.}
\label{Tatm}
\end{table*}
 
\begin{table*}[ht]
\centering
\begin{tabular}{cccccccc}
Pass-band & P$_{\rm scope}$ & P$_{\rm atm}$ & P$_{\rm e}$ & & BLIP & Detector& Measure
d\\
\ [GHz] & \multicolumn{3}{c}{[pW]} & & \multicolumn{3}{c}{NEP [$10^{-17}\,{\rm W}\,{\rm Hz}^{-1/2}$]}\\
\tableline
$142$ & $10.8$ & $2.2$ & $75.0$ &&  $4.9$ & $14.7$ & $29.8$\\
$217$ & $26.3$ & $6.7$ & $87.7$  && $9.7$ & $16.7$ & $72.1$\\
$269$ & $22.2$ & $7.7$ & $123.0$ && $10.4$ & $18.0$ & $125.2$\\
\end{tabular}
\caption[]
{Background power and sensitivity.
$P_e$ is bias power dissipated in the thermistor,
$P_{scope}$ is emission from warm telescope,
and $P_{atm}$ is atmospheric emission.
BLIP is the noise contribution to a single detector due to the
statistical arrival of background photons.
Detector NEP includes contributions due to Johnson, phonon and amplifier
noise and is listed is for a single detector.
Measured NEP is from telescope scans and includes the contribution
of the atmosphere (measured at $125\,$mHz).}
\label{bolosens}
\end{table*}

The thermistor is constructed from a 
$(200\,\mu{\rm m})^2 \times 250\,\mu{\rm m}$ chip of nuclear 
transmutation doped (NTD) Ge:Ga (\cite{Haller}).
The resistance of the thermistor as a function of temperature can be 
expressed as
\begin{equation}
R(T) = R_0\exp{\sqrt{\D \over {T}}}\,,
\end{equation}
where $R_0$ and $\D$ are properties of the thermistor and $T$ is 
its temperature.
$\D$ is determined by the doping of the NTD, which is extremely uniform 
for the single batch from which all the thermistors are constructed.
$R_0$ is determined by the geometry of the thermistor
but is also effected by mechanical stress.
In order to mechanically support and provide electrical and thermal
contact to the thermistor 
without stressing it, gold wires $20\,\mu$m in diameter are 
wedge-bonded to the gold contacts of the thermistor.
These wires are attached to the bare side 
of the absorber with epoxy.
The large heat capacity of the gold leads dominate the heat capacity of
the bolometers.
The measured thermal time constants of 
$\tau \sim 50-100\,$ms easily satisfy the modest requirements of our 
drift scanning (slow modulation) observations.

Superconducting niobium-titanium (NbTi) wires connect the gold wires to 
thermally sunk electrical contacts on the bolometer mounting ring.
These $12\,\mu$m diameter leads are copper clad at the ends, where they
are indium soldered to the gold leads and lead-tin soldered to the
mounting ring electrical contacts. 
The portions of the NbTi wires with no copper cladding are 
$\approx 4\,$mm long; together the two leads contribute a thermal 
conductivity, $G \approx 10^{-10}\,{\rm W}{\rm K}^{-1}$.
The bolometer is connected to the rest of the electrical circuit by
$125\,\mu$m Manganin wires.

Fine nylon threads $\sim 15\,\mu$m in diameter (12 denier 8 filament
Nylon type 782 bright, E.~I.~Du~Pont de Nemours and Company) support
the absorber and provide a
thermal conductivity of $G \approx 10^{-9}\,{\rm W}{\rm K}^{-1}$ at 
the operating temperature of the bolometers.
For optimal performance, each spectral band could have a set of detectors
with the support structure optimized for the measured system loading.
Instead, we compromise and choose the thermal conductivity so that the
same set of detectors can be used under the range of background powers
encountered in all three spectral bands.
The measured detector noise equivalent power is
$NEP \approx 1.2\times 10^{-16}\,{\rm W}\,{\rm Hz}^{-1/2}$ 
at optimum bias (in the dark).
This noise level is comparable to the background photon
noise in the $217$ and $268\,$GHz bands 
(Table~\ref{bolosens}).

\begin{figure}[htbp]
\plotone{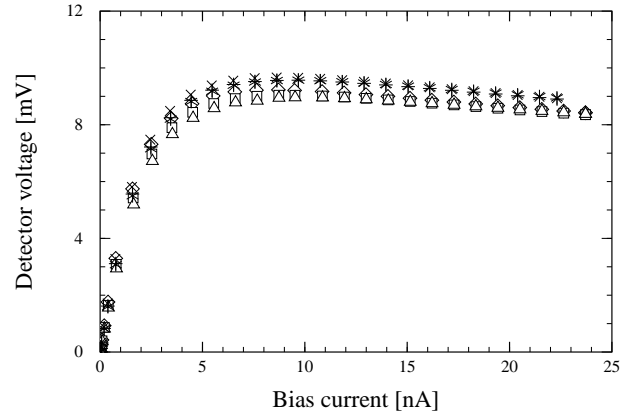}
\caption[Matching of bolometer I vs. V characteristics]
{Plot of six bolometer I-V curves.  The non-ohmic behavior, shown
in the roll-over at $\sim 5 {\rm nA}$, is caused by electrical
self-heating in the detector.  The excellent matching of the
detectors is evident, especially at high bias.
The six detectors cluster into two rows of three in which the
voltages are virtually identical.}
\label{bolomatch}
\end{figure}

The detectors must be well matched to provide acceptable 
performance in the bridge readout (Section~\ref{acbridge}).  
In order to provide a large common mode rejection ratio (CMRR) 
to atmospheric noise, the thermal conductivity of the
detectors should be matched to better than $5\%$, while the parameters
describing the thermistor resistance, $R_0$ and $\D$, should be matched 
to better than $5\%$ and $1\%$, respectively (\cite{Glezer}).  
Matching of thermistor characteristics is ensured by the
uniform doping of the NTD Ge:Ga material and careful control of the
geometry and contacting of the thermistors.  
The matching of the detector thermal conductivities 
is controlled by the use of uniform materials and strictly
defined geometries for the substrate and mounting. 
The values of $R_0$, $\D$, and $G$ 
for the six bolometers are given in Table~\ref{bolotab}.
The measured dispersion of the values of $R_0$, $\D$, and $G$ 
about their mean are $4.8\%$, $0.8\%$, and $5.2\%$, respectively.
When biased identically, the detectors within a given row have
output voltages matched to a within $\approx 1\%$ (Fig.~\ref{bolomatch}).  
During observations, when the bias currents are adjusted to
match the responsivities of the detectors to common mode 
atmospheric emission, the output voltages of the detectors 
match within $\approx 1.5\%$.
To improve the detector matching and maintain a high CMRR in the 
presence of changing background power,
the detectors are usually biased with a large current, 
$10-15\,$nA, this decreases the sensitivity 
to $NEP \approx 1.5-2.0 \times 10^{-16}\,{\rm W}\,{\rm Hz}^{-1/2}$.
As is shown in Table~\ref{bolosens}, the system noise is dominated 
by atmospheric noise; 
a small decrease in detector sensitivity has little effect on the 
performance of the instrument.

\begin{table}[ht]
\begin{center}
\begin{tabular}{cccc}
Channel & $R_0(\Omega)$ & $\D$ & $G(pW/K)$\\ \tableline
$1$ & $13.9$ & $51.2$ & $1015$\\
$2$ & $14.6$ & $51.6$ & $955$\\
$3$ & $14.3$ & $51.0$ & $1055$\\
$4$ & $15.4$ & $50.7$ & $946$\\
$5$ & $15.6$ & $51.6$ & $878$\\
$6$ & $14.0$ & $51.3$ & $976$\\ \tableline
Average & $14.6$ & $51.2$ & $963$\\
$\sigma$ & $0.7$ & $0.4$ & $50$\\
\end{tabular}
\end{center}
\caption[]
{Bolometer characteristics. 
The thermal conductivity is measured in the dark at the operational
temperature of the bolometers in the $142\,$GHz band.}
\label{bolotab}
\end{table}
 
\subsection{AC Bridge}
\label{acbridge}
The major technical advance represented in this instrument lies in the
low-frequency stability of the bolometer difference signals.
This low-frequency stability, provided by a balanced AC bridge readout, 
allows drift-scan observations to be made with no mechanical modulation 
of the beam and with the telescope fixed relative to the Earth.  
In this observation mode, only
celestial sources of emission are modulated.  The ratio of time on and
off source is comparable to that for a multi-level
differencing scheme (\cite{Herbig}).

\begin{figure}[htbp]
\plotone{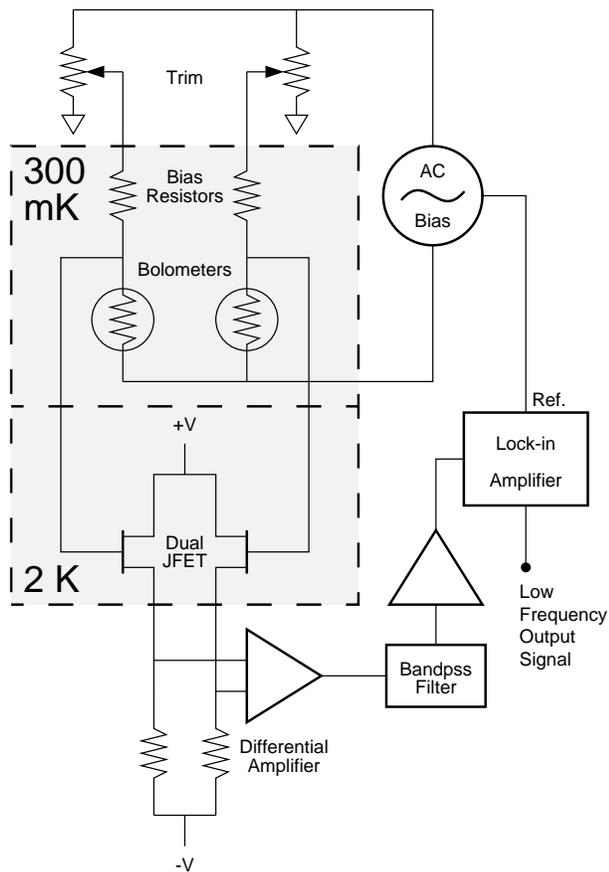}
\caption[AC-bridge readout for bolometers]
{AC-bridge readout for bolometers.  The bias resistors are large
compared to the trim resistors, which are needed to compensate
for mismatches in the bolometer responsivities.  The initial
amplification stage is mounted on the $^4$He cold-plate to reduce
the amount of high impedance wiring.  In the SuZIE experiment,
both bolometers are exposed to the sky, and the output is
proportional to the differential sky brightness.}
\label{bridge}
\end{figure}

Each AC-bridge contains a pair of matched bolometric detectors and
bias resistors as shown in Figure~\ref{bridge}
(\cite{Rieke}; \cite{Wilbanks}; \cite{Devlin}).  
The bolometers register a small change in 
optical power as a proportional change in the resistance of the 
thermistor.
The circuit described here produces a stable voltage proportional
to the difference in resistance between two detectors.
We use this circuit in
the SuZIE receiver to difference between pixels
in the focal plane, allowing us to determine the differential 
brightness of two points on the sky without modulation of the incoming 
beam.  
In principle (for perfectly matched detectors and optics), this 
readout system perfectly suppresses signals due to
fluctuations in bias amplitude, amplifier gain,
heat-sink temperature, and common mode atmospheric noise.

This system is still susceptible, as are chopping systems, 
to noise due to differential atmospheric emission.  
In this regard, the differencing is equivalent to square-wave 
beam-switching at infinite frequency.
True differential noise arises because the six beams 
in the array are each sensitive to emission from slightly
different columns of atmosphere.
Adjusting the bias current of a detector changes
its voltage responsivity to absorbed power; this allows the responsivities 
of the detectors to be accurately matched.  
Glezer \ea (1992) showed that, in general, 
it is possible 
to null the response of the detector differences to common-mode optical 
signals for reasonable ratios of electrical bias power to incident 
optical power.  
They also point out that the CMRR,
in the presence of changing optical loading, is 
improved by increasing the detector bias power beyond the point 
where maximum signal to noise is obtained in the
absence of common mode optical signals.

\begin{figure}[htbp]
\epsfxsize 3.0 in
\epsfbox{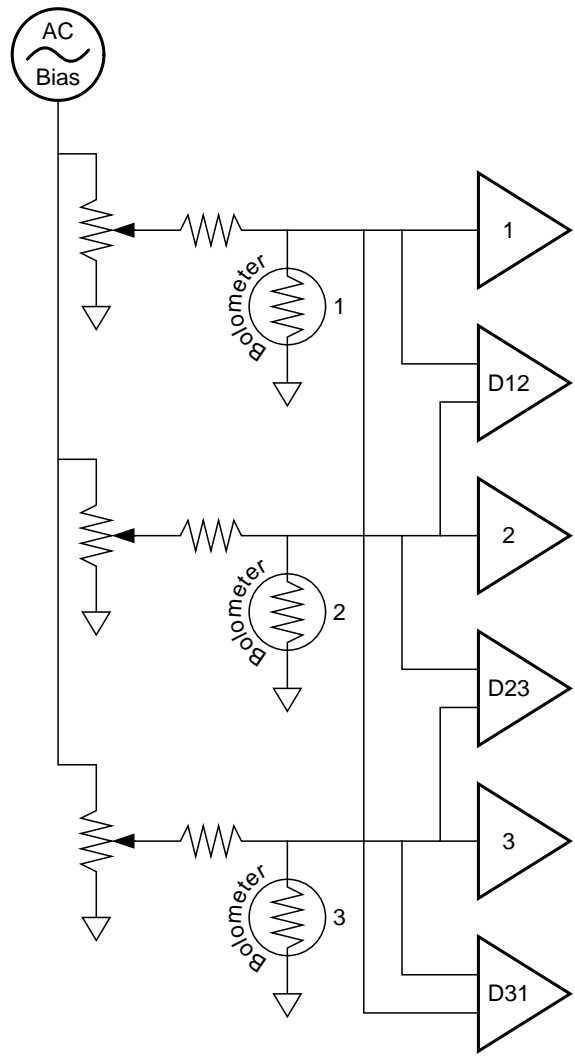}
\caption[SuZIE readout electronics]
{One row of the SuZIE readout electronics.
Each row has a single bias, three difference
channels, and three absolute channels.}
\label{readout}
\end{figure}

\subsection{Read Out Electronics}
\label{sreadout}
The SuZIE detector electronics, shown in Fig.~\ref{readout}, are
configured to take differences only along the long direction of the
array.  Among the three detectors in a row, the electronics read out
all three possible differences, giving a total of six for the array.

The voltages of the detectors in each bridge are buffered by a 
matched pair of cold JFET (NJ132L, InterFET Corp.) source followers 
(six pairs total for the six differences). 
The JFETs are mounted on the $2\,$K cold-plate and connected to the 
$300\,$mK stage by short, tightly constrained Manganin leads.  
The proximity of this first amplification
stage to the detectors isolates the high-impedance points in the
circuitry from the wiring out of the cryostat, minimizing the amount
of wiring susceptible to microphonic pickup, channel-to-channel crosstalk,
and radio frequency interference.  
The JFETs
are thermally isolated from the $2\,$K cold-plate on fine fiber-glass
stalks to allow them to maintain an operating temperature near $100\,$K.
They are encased in $2\,$K copper cavities to prevent their thermal
radiation from reaching other parts of the instrument.
At the operating frequency of $\sim 100\,$Hz, the JFET source followers
exibit no $1/f$ noise and each 
contributes $\approx 2.0\,{\rm nV}\,{\rm Hz}^{-1/2}$ 
noise to the signal it buffers.
\begin{figure}[hbtp]
\plotone{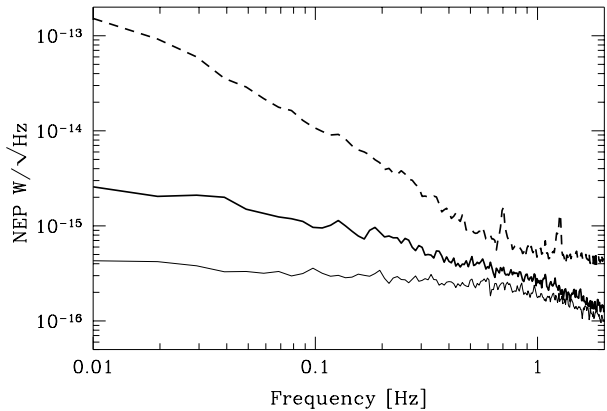}
\caption[SUZIE system noise]
{SuZIE system noise for a detector difference as measured
in the laboratory with an optical blank in place of the
Lyot stop (solid line).
Also shown is the noise for the system at the telescope in the
$217\,$GHz configuration, for both a single detector (dashed heavy line)
and for a detector difference (heavy line).
The results are calibrated in terms of NEP using the measured
electrical responsivity of the detectors.}
\label{noise}
\end{figure}
The bridge is biased with a square-wave bias rolled off at frequencies
above the third harmonic.
All three detectors in a row are driven by a common bias; the two 
rows operate at different frequencies, both near $100\,$Hz. 
At this frequency, the amplifiers have low noise and the bias 
oscillations are much faster than the bolometer time constants.
The buffered detector signals are subtracted in a low noise 
differential amplifier and the difference signal is demodulated 
in a square-wave lock-in amplifier.
A filter, similar to that used to roll-off the bias, is placed 
before the input of the lock-in to reduce the sensitivity of the 
demodulated signal to noise near higher harmonics of the 
fundamental.
The outputs of the
demodulators pass through 2-pole low-pass Bessel filters set at
$2.25\,$Hz, just less than half the $5\,$Hz A/D sampling frequency.
Outside the cryostat electronics box, another amplification stage
removes DC offsets from each channel and matches the dynamic 
range of the signals to the analog-to-digital converter.  
This gain stage includes an additional set of 
2-pole anti-aliasing low-pass Bessel filters set to $2.25\,$Hz.
Six similar circuits read out the voltage on each bolometer
individually.  These are essential for characterizing the behavior of
the instrument and the detectors.
They also provide a measure of sky
brightness over time.  For these circuits, the signals pass through 
AC-coupled amplifiers before the demodulators.  
The gain of the single detector readouts is $500$  compared 
with $7 \times 10^4$ for the difference channels.
In Fig.~\ref{noise} we show the noise, normalized to unity gain, 
for a single detector and the $4.3^{\prime}$ difference viewing the
sky in the $217\,$GHz band.
The system noise in both the single detectors and differences
is dominated by atmospheric emission.
Between $10$ and $100\,$mHz (the signal bandwidth for drift scans 
over a cluster) the 
CMRR to atmospheric noise ranges from $\sim 60$ to $10$. 

The 12 bolometer signals, two channels monitoring the amplitude of the
bias voltages, and two channels that monitor the temperature control
circuitry (see section~\ref{scryo}) are read by a 16 channel, 16-bit A/D
converter (Model 576 High Speed Data Logging System, Keithley
Instruments, Inc.) and transferred to an Apple Macintosh IIci computer
using IEEE-488 communications.

\subsection{Parallactic Alignment Rotator}
\label{srot}
The SuZIE system is designed specifically for drift-scan observations 
made with the telescope immobilized.  
In this scheme, the telescope is pointed
west of the source and fixed in altitude and azimuth.  The
earth's rotation then sweeps the telescope beam over the source.  In
order to orient the array with the rows aligned East/West so that the
source drifts sequentially through each of the three pixels in a row, 
the entire receiver is mounted in bearings that rotate about the optical
axis.  At $0^{\circ}$ rotation angle the array is parallel to the
horizon, appropriate for observing sources at transit.  The receiver
can be rotated by angles up to $90^{\circ}$ in either direction from 
this position to allow observation of sources rising to the east
or setting to the west of the telescope.

A computer controlled stepper motor keeps the array aligned with the
scan direction, updating the rotation angle at the beginning of every
scan ($\approx 2$ minutes).  
A precision ten-turn variable resistor is engaged to the belt
that turns the receiver.  
Its resistance is calibrated and then monitored to ensure the 
integrity of the mechanical linkage from the stepper motor to the 
receiver.
\subsection{Cryogenics}
\label{scryo}
The SuZIE receiver is built into a liquid helium cryostat with a
liquid nitrogen cooled intermediate heat shield (Infrared
Laboratories).  The L$^4$He and LN$_2$ tanks have $4.5$ liter and
$3$ liter capacities, respectively.  
During operation, the L$^4$He bath is
pumped to achieve cold-plate temperatures below $2\,$K.  Both the
concentrators and the detectors are enclosed within the L$^4$He cooled
heat shield.  They are thermally isolated from the L$^4$He cold-plate
on Vespel (E.~I.~Du~Pont de Nemours and Company) tubes and cooled to
$300\,$mK by a $^3$He refrigerator.
The $^3$He refrigerator is a self-contained charcoal sorption pump
refrigerator, similar to those described by
Duband \ea (1991).
The refrigerator holds $4\,$STP liters of $^3$He gas at
$6-7\,$MPa at room temperature, and has a hold-time of more than
12 hours.  The operating temperature of the detector cold-stage is
$\sim 305\,$mK under the heat load in the receiver; it is raised to
$314\,$mK in order to facilitate temperature control.
\begin{figure}[htbp]
\plotone{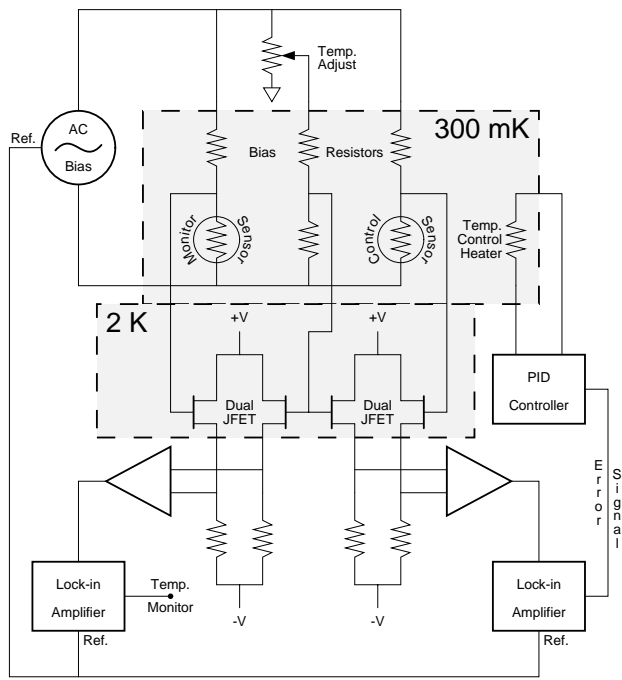}
\caption[SuZIE temperature control electronics]
{The SuZIE temperature control electronics.  Two thermometers mounted
on the optics block are differenced against fixed resistors.
A variable bias resistor allows adjustment of the temperature set
point.  The output from one bridge is used as the error signal for a
proportional-integral-differential (PID) controller, which closes the
control loop using a heater on the optics block.  The other bridge is
monitored to determine the temperature stability.}
\label{tempcont}
\end{figure}

\begin{figure}[htbp]
\plotone{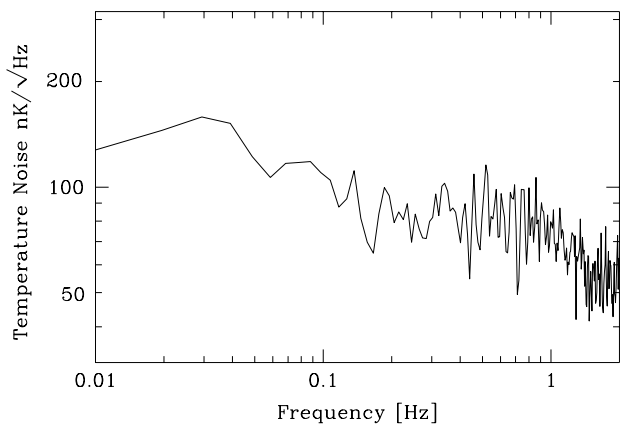}
\caption[Measured temperature noise]
{Measured temperature noise of the $300\,$mK cold stage monitor thermistor.}
\label{templot}
\end{figure}
 
Initial temperature control using commercially available instruments
was unsatisfactory and a customized precision temperature control 
system was developed.  Although
the commercial sensor (GRT-200, Lakeshore Cryotronics) and resistance
bridge thermometer readout (AVS-46 Resistance Bridge, RV-Elektroniikka
Oy) were retained to monitor the absolute temperature, additional
sensors and readouts were installed for the control circuit
(Fig.~\ref{tempcont}).  Two NTD Ge:Ga thermistors are mounted, via
gold electrical leads, to heat-sinks on the detector cold-stage.
The two sensors operate independently, one to control the temperature
and the other to monitor it's value.
The thermistors are read out in bridge circuits identical to the
bolometer readout circuits, except the thermistors are differenced
with a fixed resistor.
The fixed resistor is chosen to be close to the resistance of the
thermistors at the chosen operating temperature.
Fine adjustment of the operating temperature is possible by 
changing the relative biasing of the resistor and thermistor.
The output of the control sensor
bridge circuit is input to a commercial PID controller (TS-530
Temperature Controller, RV-Elektroniikka Oy) which feeds back to a
$500\,\Omega$ heater resistor on the detector cold-stage.
At the operating temperature of $314\,$mK, the heater supplies 
$5-10\,\mu$W of power to the cold-stage.
The output of both the control and monitor sensor
circuits are amplified and logged as for the detector differences.

The PID controller is operated to keep the control
thermistor-resistor bridge output equal to zero. 
The monitor circuit then provides a measurement of the 
difference in temperature (if any) between the two thermistors 
and an estimate of the temperature fluctuations 
in case the control thermistor malfunctions.
In Figure~\ref{templot}, we show the Power Spectral Density (PSD) 
of the temperature 
fluctuations measured by the monitor circuit.
The temperature of the $300\,$mK cold stage fluctuates by
$< 150\,{\rm nK}\,{\rm Hz}^{-1/2}$ on time-scales of $100\,$s. 
These fluctuations make a small contribution to the total 
noise of a single detector, 
$NEP_T< 5.4 \times 10^{-17}\,{\rm W}\,{\rm K}^{-1}$.
The bolometer differences exhibit measured CMRRs to temperature 
fluctuations ranging from $12$ to $360$.
The contribution of the measured temperature fluctuations to the
detector noise is  
$NEP_T < 3.4 \times 10^{-18}\,{\rm W}\,{\rm K}^{-1}$ 
for the most poorly matched detector pair.

In operation at the telescope, there is a small signal in the 
temperature monitor at the beginning of the scan due to the 
telescope motion heating the cold stage by $\lesssim 100\,$nK. 
The cold stage recovers from this impulse with a time 
constant of $\sim 7\,$s.
Although not visible in a single scan, this feature is
significant in the coadded temperature monitor scans.
In Church \ea (1996) the effect of this
signal on the data channels is investigated 
by correlating the co-added differential signals to the temperature 
monitor signal; no significant correlation is found. 

\section{Beam-shapes and Calibration}
\label{ccal}
The instrument makes a differential measurement of the sky brightness
between pairs of array elements.  Accurate calibration requires a
knowledge of the spectral response (see Sec.~\ref{soptics}), the shape
of the beams on the sky, and the responsivity of the array.

\begin{figure}[htbp]
\plotone{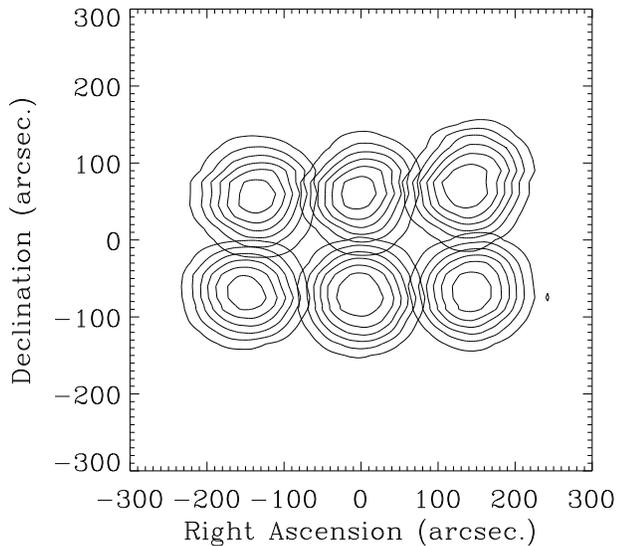}
\caption[SuZIE beam map]
{Map of bolometer single channels at $142\,$GHz constructed from drift
scans across Jupiter spaced by $15^{\prime \prime}$ in declination.
Contours correspond to 0.9, 0.75, 0.6, 0.45, 0.3 and 0.15 of maximum.
The large size of Jupiter ($\sim 40^{\prime \prime}$) broadens the beams
slightly.}
\label{beamap}
\end{figure}

The shapes of the beams are determined by drift-scan observations of
planetary calibrators. 
A map of the single-detector beam patterns measured at $142\,$GHz 
with drift scans across Jupiter is shown in Figure~\ref{beamap}. 
Scans were performed with
the array offset in declination from the source (transverse to the
scan direction) by steps of $15^{\prime \prime}$. 
The angular size of Jupiter ($\sim 40^{\prime \prime}$) broadens 
the measured beams slightly.
The beams are separated by $2.2^{\prime}$ (adjacent) and 
$4.3^{\prime}$ (end-to-end) along the direction of the rows.
\begin{figure}[htbp]
\plotone{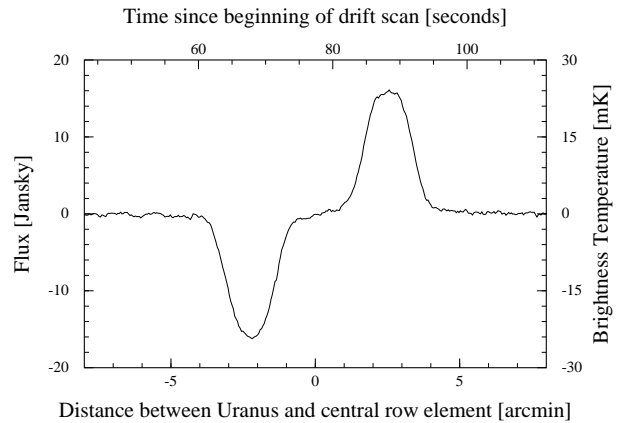}
\caption[Uranus drift scan]
{$142\,$GHz band response of $4.3^{\prime}$ difference drift
scanned across Uranus.
The brightness temperature axis corresponds to a source that fills the
$\sim 3\,{\rm arcminute}^2$ beam.}
\label{beam}
\end{figure}
The full width at half maximum (FWHM) and integrated solid angle
of each array element, measured for each of the three spectral 
bands, are listed in Table~\ref{beamstat}.
The planets used in these measurements, Mars and Uranus, had angular
diameters $\sim 3^{\prime \prime}$ and $\sim 2^{\prime \prime}$, small 
enough to be neglected in the determination of the beam parameters. 
The parameters listed for the $142\,$GHz and $217\,$GHz bands were
measured after the installation of a new tertiary mirror, which
significantly improved the measured beam patterns. 
The improved matching of the beams is evident from the reduced scatter 
when compared to the $268\,$GHz band beams measured in 1993 with 
the original tertiary mirror.
After the installation of the new tertiary mirror, the measured FWHM for
the pixels in the array match to within the accuracy of our measurement
$(\approx .05^{\prime})$.

\begin{table*}[htb]
\begin{center}
\begin{tabular}{ccccccc}
Pass-band & \multicolumn{2}{c}{$142\,$GHz} & \multicolumn{2}{c}{$217\,$GHz} &
 \multicolumn{2}{c}{$269\,$GHz*}\\
Channel& FWHM (arcmin) & $\Omega$ (arcmin$^2$) & FWHM (arcmin) & $\Omega$ (arcmin$^2$) & FWHM (arcmin) & $\Omega$ (arcmin$^2$)\\ \hline
$1$ & $1.78$ & $3.25$ & $1.93$ & $3.75$ & $1.57$ & $2.39$\\
$2$ & $1.83$ & $3.19$ & $1.99$ & $4.19$ & $1.70$ & $2.82$\\
$3$ & $1.77$ & $3.16$ & $1.97$ & $3.74$ & $1.92$ & $3.44$\\
$4$ & $1.74$ & $3.11$ & $1.87$ & $3.46$ & $1.91$ & $3.45$\\
$5$ & $1.68$ & $2.96$ & $1.88$ & $3.71$ & $1.78$ & $3.20$\\
$6$ & $1.80$ & $3.42$ & $1.91$ & $3.46$ & $1.60$ & $2.50$\\ \hline
Average & $1.77$ & $3.18$ & $1.92$ & $3.72$ & $1.75$ & $2.97$\\
$\sigma$ & $0.05$ & $0.15$ & $0.05$ & $0.27$ & $0.15$ & $0.47$\\
\end{tabular}
\end{center}
\caption[Width and Solid Angle of Beams]
{Full Width Half Maximum (FWHM) and solid-angle
$(\Omega$) of the beams for each array element,
as determined from scans of Mars and Uranus.
 
(*) The $269\,$GHz beams were measured with the original tertiary mirror.}
\label{beamstat}
\end{table*}

The planet scans provide the absolute calibration of the
instrument.  The output signals are divided by the known source 
brightness, yielding the responsivity to celestial sources.  
The observations at $142$ and $217\,$GHz (April 93; April 
and May, 1994) were calibrated by comparing scans over Uranus 
(Figure~\ref{beam}) with the brightness model of 
Griffin and Orton (1993) integrated over 
the measured spectral bands.
The $268\,$GHz observations (May 1993) were calibrated with Mars 
assuming a brightness temperature of $195.6\,$K  (\cite{Orton}).

\section{Scan Strategy}
We have developed a scan strategy that minimizes and
provides sensitive tests for systematic errors.
The telescope acquires a position leading the source by a
right ascension offset (RAO) and tracks that position. 
The array is rotated such that the rows of detectors lie along lines of 
constant declination.
To begin the scan, the tracking is disabled and the telescope 
maintains a fixed altitude and azimuth.
The rotation of the earth then moves the beams in increasing RA 
at a rate of $.25\, {\rm cos}(\delta)\,^{\prime}/{\rm s}$.
During the scan, there are no moving parts and only
celestial sources are modulated. 
After drifting for $\sim 30^{\prime}$ the scan ends and the telescope
re-acquires the position leading the source.
During all observations, we alternate the RAO of sequential 
scans between $12^{\prime}$ and $18^{\prime}$.

Because of the low surface brightness of the S-Z effect, we
have to combine many scans in order to produce a significant 
detection.
Our results are, therefore, sensitive to the presence of any scan 
correlated instrumental baseline.
We check for such a instrumental baseline in two
ways.

First, we observe patches of sky free of known sources.
We analyze the data in the same way as the source data. 
An instrumental baseline would be detected as a non-zero signal
when the source model is fit to the scan data. 
In 1994, we accumulated 8.5 hours of data on one patch
of sky centered at $(10^{\rm h}\,24^{\rm m}\,25^{\rm s},\; 
3^\circ\,49\pr\,9\2pr\; {\rm J}2000)$, 
and 7.5 hours on a second patch centered at 
$(16^{\rm h}\,32^{\rm m}\,44^{\rm s},\; 5^\circ\,49\pr\,43\2pr\; {\rm J}2000)$.
Fitting this data to an isothermal source model described by 
$(T_e=10\,{\rm keV},\;\beta = 2/3,\;\theta_c = 1.2^{\prime})$, we find 
a central Comptonization $y_0 = 1.5 \pm 2.2 \times 10^{-5}$.
Therefore, to this level of sensitivity, sufficient for
observations of a large sample of clusters, we find no 
significant baseline.
We have also determined the instrumental baseline when
observing in the $1.4$ and $1.1\,$mm bands.
We find no significant baseline; however, because of the 
much larger sky noise, the limits are less sensitive.

Although a baseline must be correlated
between several scans to be significant, it could vary on 
long time-scales.
For this reason, it is preferable to check each source 
observation for the presence of an instrumental baseline.
In the observations, the RAO is alternated between scans 
in order to make such a test possible.
Alternating the RAO moves the source by $\D {\rm RAO} = 6^{\prime}$
while any baseline is unaffected.
Subtracting sequential pairs of scans then removes the 
baseline with only a small effect on the signal.
The subtracted scans are fit with the difference of two 
source models offset by $\D {\rm RAO}$.
In Holzapfel \ea (1997a) this method is used
to test for the presence of an instrumental baseline in the 
$142\,$GHz data for A2163.
In all cases, there is no significant change in the results
when the scans are differenced, indicating that there is no 
significant instrumental baseline. 
\section{Sensitivity}
\label{sense}
\begin{figure}[tbp]
\plotone{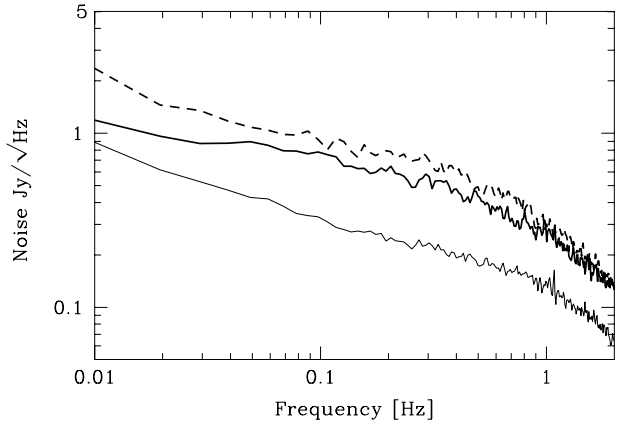}
\caption[Noise in SuZIE bands]
{SuZIE system noise for the difference between 2 detectors separated
by $4.3^{\prime}$.
The sold line corresponds to the $142\,$GHz band, the heavy line
the $217\,$GHz band, and the dashed line the $268\,$GHz band.
The data for three spectral bands are calibrated in terms of Janskys
by observations of planets.}
\label{atmnoise}
\end{figure}

The sensitivity of the system to point sources is determined from
the measured responsivity and noise.
Fig.~\ref{atmnoise} shows the PSD of the noise for the same 
$4.3^{\prime}$ difference during observations in the $142$, 
$217$, and $268\,$GHz configurations. 
The noise performance at frequencies below $300\,$mHz is dominated by noise 
resulting from differential atmospheric emission.
The atmospheric noise varies
considerably from one observation to another; the data shown here are 
chosen to be representative of the performance in good weather.
At frequencies of $\sim 125\,$mHz the sensitivities in the $4.3^{\prime}$
difference are approximately $300\,{\rm mJy\,Hz}^{-1/2}$, $800\, 
{\rm mJy\,Hz}^{-1/2}$, and $1000\,{\rm mJy\,Hz}^{-1/2}$ for 
observations in the $142$, $217$, and $268\,$GHz bands. 

\begin{table*}[htb]
\begin{center}
\begin{tabular}{ccccc}
Pass-band & $\D y$ & $\D v_r \times (0.01/\tau)$ & $\D T_{RJ}$ & $\D T_{CMB}$
\\
$[{\rm GHz}]$ & $10^{-5}\sqrt{\rm Hr}$ & $[{\rm kms}^{-1} \sqrt{\rm Hr}]$ &
 $[{\rm mK}\sqrt{\rm s}]$ & $[{\rm mK}\sqrt{\rm s}]$\\ \hline
$142\,$GHz & $6.8$ & $2010$ & $1.3$ & $2.1$\\
$217\,$GHz & $-$ & $3190$ &  $0.9$ & $3.5$\\
$269\,$GHz & $29$ & $5500$ & $1.1$ & $5.7$\\
\end{tabular}
\end{center}
\caption[Sensitivity to S-Z effects and $\D T$]
{Typical sensitivity to $y$ and $v_r$ for a cluster with isothermal
IC gas described by
$(T_e=10\,{\rm keV},\, \theta_c=1.2^{\prime},\,{\rm and}\, \beta = 2/3$).
Also listed are the raw sensitivities for a signal
that completely fills one beam of the difference, where
$\D T_{RJ}$ and $\D T_{CMB}$ correspond to changes in
temperatures of Raleigh-Jeans and CMB sources.}
\label{senstab}
\end{table*}

We have used observations of S-Z sources to characterize the sensitivity 
of the system to central Comptonization and peculiar velocity.
An isothermal model for the surface brightness is computed 
assuming the IC gas to be described by 
$(T_e = 10\,{\rm keV},\; \theta_c= 1.2^{\prime},\; \beta =2/3)$.  
Relativistic corrections have been taken into account; they
decrease the sensitivity to $y$ slightly.
The source model is convolved with the measured beam patterns in order 
to determine a template for the bolometer difference signals during
the drift scan.
All six bolometer differences are fit to the source template
in order to determine the source amplitude for each scan.
The uncertainty is determined from the scatter of the values obtained
from each scan about their weighted mean.
The sensitivities achieved for the three spectral bands are 
listed in Table~\ref{senstab}; these are the results of 
$5-7$ hours of observation taken on each of three separate nights. 
The atmospheric noise is extremely variable, resulting in values of 
noise as much as two times larger than these values on some nights.
The atmospheric noise is correlated with the zenith optical depth 
$\tau(225\,{\rm GHz})$,
however, it is more strongly correlated with the instability
of $\tau(225\,{\rm GHz})$ over the course of an evening.  

We have also determined the sensitivity of the SuZIE system
to changes in the temperature of Rayleigh-Jeans and CMB sources.
We assume that one beam of a detector difference 
is completely filled by the perturbed temperature.
The noise used in the calculation is for $\sim 125\,$mHz 
which corresponds to the time spent on a $\sim 2^{\prime}$ 
beam when drift scanning.
The resultant sensitivities for the three spectral bands are
listed in Table~\ref{senstab}.

\section{Applications to Cosmology}
\label{cappl}
\subsection{Hubble Constant}
Using the SuZIE system, we have detected decrements in
the brightness of the CMB in the direction of the clusters A2163, 
A1689, A1835, A2204 and ZW3146 with significance greater than $6\,\sigma$.
In Holzapfel \ea (1997b), we combine observations of 
the S-Z effect in A2163 with X-ray data
in order to determine the Hubble constant.
Assuming the IC gas to be isothermal, and including
many possible sources of systematic error, we find
$H_0= 60^{+45}_{-31}\,{\rm kms}^{-1}$ at $68\%$ confidence, where the 
uncertainty is dominated by the IC gas temperature.
Adopting the thermal structure determined by ASCA (\cite{Markevitch}), 
$H_0= 78^{+60}_{-40}\,{\rm kms}^{-1}$.

The uncertainty in the distribution of the IC gas along the line of 
sight limits the accuracy of a Hubble constant determined
from only one cluster to $\sim \pm 27\%$.
In order to reduce this uncertainty we must average the results of 
a carefully selected sample of clusters while taking care
to control systematic errors.
The uncertainty in the ellipticity will dominate if we are able to 
to determine $y$ to approximately $15\%$ accuracy.
In the ROSAT all sky survey (\cite{Ebeling}) there are $\sim 25$ clusters 
at redshift $z>0.2$ with 
$L_X(0.1-2.4\,{\rm keV}) > 10^{45}\, {\rm erg\,s}^{-1}$;
we estimate these clusters to have $y \greatsim 1.5 \times 10^{-4}$.
In $\sim 10$ hours per cluster, the SuZIE instrument can make
observations of the S-Z effect which, assuming high quality X-ray 
data, can be used to determine 
$H_0$ with uncertainty limited by cluster ellipticity.
 
\subsection{Peculiar Velocities}
Galaxy clusters have been shown to be efficient tracers of the
large-scale velocity field of the universe
(\cite{Bahcall}; \cite{Gramann}).
We have used measurements at mm wavelengths to constrain peculiar 
velocities of the clusters A1689 and A2163.
Unlike the determination of the Hubble constant, the details
of the IC gas distribution are not important to the determination of the
peculiar velocity.
The resulting peculiar velocities depend weakly on the IC gas 
temperature.
The uncertainty is dominated by the statistical uncertainty in the 
amplitude of the S-Z kinematic component.
The combination of $142$, $217$, and $268\,$GHz measurements have been 
used to place limits on the radial component of peculiar velocity for 
A2163 of $v_r=+490^{+1370}_{-880}\,{\rm kms}^{-1}$ at $68\%$ confidence.
The combination of $142$ and $217\,$GHz observations of A1689
have been used to determine 
$v_r=+170^{+815}_{-630}\,{\rm kms}^{-1}$ (\cite{Holzapfelb}).

\subsection{Primary Anisotropies and Source Counts}
Although specifically designed to observe the S-Z effect in X-ray
selected clusters, the SuZIE receiver is a powerful tool for 
measuring other anisotropies of the CMB.

In April 1994 we observed, at $142\,$GHz, two patches of sky, each 
approximately $36^{\prime} \times 4^{\prime}$ and free of known sources.
In the $\sim 14$ hours spent on both patches we reached a flux limit
of $\approx 10\,$mJy for $80$ $1.7^{\prime}$ FWHM pixels. 
Assuming a Guassian autocorrelation function for the CMB fluctuations, 
Church \ea (1996) set a $95\%$ confidence upper 
limit of $\D T/T \leq 2.1 \times 10^{-5}$
for a coherence angle of $1.1^{\prime}$.
This result is comparable to limits set by other arcminute scale 
anisotropy
experiments, but is less prone to confusion by radio sources.
We anticipate significant increases in both the sensitivity and sky 
coverage in the near future.

The SuZIE instrument has also been used to investigate 
structure detected by 
degree scale CMB anisotropy experiments.
During the April 1994 observing run, we scanned two 
$1^\circ \times 1^\circ$ 
fields searching for unresolved sources that may be contributing to
structure in degree scale anisotropy measurements
(\cite{MSAM}; \cite{MAXGUM}).  These surveys were performed by 
slewing the telescope in azimuth at $1^{\prime}/{\rm s}$.  
The observation of each field took $\sim 1.5$ hours.  
Depending on the atmospheric noise in these
low-elevation angle ($\sim 20^\circ$) observations, $3\,\sigma$ limits
between $0.5$ and $1.0\,$Jy on the brightness of sources less than
$2^{\prime}$ in extent were achieved over the whole 
field (\cite{Church95}).

The surveying capabilities of the SuZIE instrument make it ideal
for searching for the S-Z effect in distant clusters.
In a low density universe, clusters cease to grow after a redshift
$z \sim (\Omega_0^{-1} - 1)^{-1}$.
Therefore, information about the number density of S-Z bright clusters at 
high redshift is a powerful probe of $\Omega_0$.
Unlike the X-ray emission from clusters which is dominated by the
emission from the central core radius, the S-Z effect depends only 
on the quantity of hot gas, not on details of its distribution. 
In $\sim 200$ hours of observation, SuZIE could be used to image
a $1^\circ \times 1^\circ$ patch of sky to a flux limit of $10\,$mJy.
At this sensitivity, between $3\ (\Omega_0 = 1.0)$ and 
$10\ (\Omega_0 =0.3)$ clusters should be detected (\cite{Markevitch};
\cite{Barbosa}; \cite{Eke}).
\section{Conclusion}
\label{cconc}
The SuZIE effort has shown that sensitive millimeter wave observations
of the S-Z effect in distant galaxy clusters can be made using
ground-based telescopes. 
Using new filter, detector, and readout technologies the
receiver described here produces
measurements of differential surface brightness stable on long
time scales.
Observing with drift scans, which minimize systematic errors, 
this instrument can be used to measure low surface brightness signals.
The sensitivity of the system is limited by residual differential 
atmospheric emission in all three frequency bands.

A second, more sensitive, array receiver is under construction.  The
new receiver will contain an array of four color
photometers, each photometer similar in design to that developed for
the Far InfraRed Photometer on the InfraRed Telescope in Space
(\cite{Lange94}).  Simultaneous observations will be made at $142$,
$217$, $268$, and $360\,$GHz in each pixel of the array.
Multifrequency observations will increase the efficiency of our 
observations, but more importantly will provide a
powerful tool for the removal of residual atmospheric noise.
The residual atmospheric noise evident in Fig.~\ref{noise} can be
removed during data reduction by exploiting the difference 
in the spectra between the atmosphere and the components of the S-Z effect. 
The removal of atmospheric noise will particularly impact the higher
frequency channels and improve the sensitivity to peculiar velocity.
With this improvement, the detection of individual cluster peculiar
velocities should be possible. 

We would like to thank Sabrina Grannan, Teresa Ho, and Ken Ganga for
their contributions to observations and data analysis.
Thanks to Antony Schinkel and the entire staff of the CSO for their
excellent support during the observations.
The CSO is operated by the California Institute of Technology under
funding from the National Science Foundation, contract \#AST-93-13929.
This work has been supported by a NSF PYI grant to A. E. Lange, 
a grant from the David and Lucile Packard foundation and by National 
Science Foundation grant \#AST-95-03226.

\markright{REFERENCES}

\end{document}